\documentclass[twocolumn,superscriptaddress,aps,prb]{revtex4-1}
\usepackage{bm}
\usepackage{amssymb,amsmath}
\usepackage{comment}    
\usepackage{times}

\usepackage[dvips]{graphicx}
\usepackage{array}

\begin{document}

\title{Localization length exponent in two models of quantum Hall plateau transitions}

\author{Qiong Zhu}
\affiliation{International Center for Quantum Materials, Peking University, Beijing
100871, China}
\affiliation{Zhejiang Institute of Modern Physics, Zhejiang University, Hangzhou 310027, China}

\author{Peng Wu}
\affiliation{Zhejiang Institute of Modern Physics, Zhejiang University, Hangzhou 310027, China}

\author{R.~N. Bhatt}
\affiliation{Department of Electrical Engineering, Princeton University, Princeton, New Jersey
08544, USA}

\author{Xin Wan}
\affiliation{Zhejiang Institute of Modern Physics, Zhejiang University, Hangzhou 310027, China}
\affiliation{Collaborative Innovation Center of Advanced Microstructures, Nanjing 210093, China}

\begin{abstract}
Motivated by the recent numerical studies on the Chalker-Coddington network model
that found a larger-than-expected critical exponent of the localization length
characterizing the integer quantum Hall plateau transitions,
we revisited the exponent calculation in the continuum model and
in the lattice model, both projected to the lowest Landau level or subband.
Combining scaling results with or without the corrections of an irrelevant length scale, we obtain $\nu = 2.48 \pm 0.02$, which is larger but still consistent with the earlier results in the two models,
unlike what was found recently in the network model.
The scaling of the total number of conducting states,
as determined by the Chern number calculation,
is accompanied by an effective irrelevant length scale exponent
$y = 4.3$ in the lattice model,
indicating that the irrelevant perturbations are insignificant
in the topology number calculation.
\end{abstract}

\date{\today}

\maketitle

\section{Introduction}
\label{sec:intro}

In the presence of a strong perpendicular magnetic field,
the Hall conductance of two-dimensional electron gases at low temperatures
is quantized to integral multiples of $e^2/h$.~\cite{Klitzing1980}
Tuning either magnetic field or Fermi energy $E_f$, one can
drive an integer quantum Hall transition from one plateau to another.
The transition is controlled by a correlation length $\xi$,
which diverges as
\begin{equation}
\xi(E) \sim |E-E_c|^{-\nu},
\end{equation}
as the Fermi energy crosses a critical energy $E_c$ (or the magnetic field $H$ crosses a critical value $H_c$, in which case $H$ replaces $E$ in the above formula).
The critical exponent $\nu$
is believed to be universal and has been studied extensively
both theoretically and experimentally.

Prior to 2009, for non-interacting electrons, the consensus~\cite{chalker88,huckestein90,Huo1992,lee93,huckestein95,kramer05,evers08}
was $\nu \approx 2.38 \pm 0.05$. This value
was also supported by experiments on samples
with short-range scatterers,~\cite{Li2005,Li2009}
even though electron-electron interactions are present
in experiments, and may be relevant.

Quite unexpectedly, when Slevin and Ohtsuki\cite{Slevin2009} redid the finite-size scaling
study of the Chalker-Coddington Network (CCN) model including corrections due to an irrelevant operator, they found
$\nu=2.593~[2.587,2.598]$, significantly larger than the previously reported values.
They found the irrelevant length scale exponent $y \approx 0.17$, which gave rise to large corrections, and altered the exponent $\nu$ significantly.
Several calculations followed. Obuse et al.\cite{Obuse2010} reported $\nu=2.55\pm0.01$ and
a larger irrelevant length exponent $y =1.29$.
Amado et al.~\cite{Amado2011}  obtained $\nu=2.616\pm0.014$ but found logarithmic
irrelevant length scale correction.
Other calculations by Dahlhaus et al.~\cite{Dahlhaus2011},
Fulga et al.~\cite{Fulga2011}, and Slevin and Ohtsuki\cite{Slevin2012}
in the CCN model also confirmed $\nu=2.56\sim 2.6$.
Obuse, Gruzberg, and Evers\cite{Obuse2012} perform a stability analysis
of the finite-size scaling of the CCN model and reported $\nu=2.62\pm0.06$,
with an irrelevant length scale correction exponent $y \geq 0.4$, which,
as the authors pointed out, is considerably larger than most recently reported values.
All these works show that with irrelevant length scale
correction included, the localization length exponent $\nu=2.55\sim2.62$.
However, the smallness of the leading irrelevant length exponent indicates that
great care is needed in numerical studies using finite-size scaling.

Despite the flurry of new and seemingly consistent results,
the correct value of the localization length exponent remains under debate.
Gruzberg et al.~\cite{gruzberg17} questioned the regular lattice setup of the CCN and
considered a general network with geometric disorder.
Numerical simulations of this new model found $\nu= 2.374 \pm 0.018$,~\cite{gruzberg17}
in agreement with the earlier results.
Bondesan et al.~\cite{bondesan17} developed an effective Gaussian
free field approach of the CCN model of
the integer quantum Hall plateau transition.
Even though the theory confirmed that the spectrum of multifractal dimensions
at the transition is parabolic, the authors warned that numerical calculations
may suffer from existence of an irrelevant perturbation which is close to
marginal (or even marginal).

Consequently, numerical calculations of $\nu$ in alternative
models~\cite{huckestein90,Huo1992,huckestein92,huckestein94,Yang1996,Wan2000,bhatt02} are desirable
to help resolve the disagreements, especially ones that are purely two-dimensional,
and do not have to rely on the crossover to one-dimension to analyze the data.
Earlier, Huo and Bhatt\cite{Huo1992} obtained $\nu=2.4 \pm 0.1$
in a two-dimensional approach using
Chern number calculations in the continuum Landau-level model.~\cite{Huo1992}
The same result was also found in the disordered Hofstadter
model.~\cite{Yang1996,Wan2000,bhatt02}
No attempt, however, has been made for the two models including corrections due to
irrelevant length scales, as done in the CCN model.

In this paper, we revisit the continuum Landau-level model and
the Hofstadter lattice model with short-range impurities.
After projecting the disorder potential into the lowest Landau level (subband),
we calculate the Chern number for all eigenstates to identify the conducting states
and perform finite-size scaling of their total number.
Combining results in the two models,
we obtain a larger exponent $\nu = 2.48 \pm 0.02$ than earlier results $\nu = 2.4 \pm 0.1$, but the two are still consistent, given the larger error bars of the previous work.
In addition, we find that the corrections to scaling are very small,
(e.g., the irrelevant length exponent is $\it large$, $y = 4.3 \pm 0.2$
in the lattice model), which can explain the relative accuracy
of the earlier results on smaller sizes using the Chern number approach, unlike transfer matrix methods.
We also perform finite-size scaling of the width of the
density of conducting states, where the results appear to be less reliable. We attribute this to the fluctuations in the tails
of the density of conducting states in finite-sized samples, which renders calculations of higher moments less reliable. Thus, the zeroth moment (the total number of conducting states) is more reliable than the width of density of conducting states, which is the second moment.

The paper is organized as follows.
In Sec.~\ref{sec:model} we introduce the two models we study.
We briefly review the Chern number calculation method
and discuss our scheme for analyzing our data. The next section describes our main results.
We first analyze the finite-size scaling of the total
number of conducting states in the lattice model and
in the continuum model in Sec.~\ref{subsec:nc_result}.
Then, we present the finite-size scaling of the width
of the density of conducting states in the two models
in Sec.~\ref{subsec:dE_result}.
We discuss the potential errors in the Chern number calculations
in Sec.~\ref{subsec:errors}.
We summarize our results and discuss our conclusions in light of other work in Sec.~\ref{sec:conclusions}.
In Appendix~\ref{app:moreLattice} we supplement our discussions with additional results in the lattice model, which, in the limit of small magnetic flux per plaquette, evolves smoothly into the continuum model.

\section{Models and Method}
\label{sec:model}
In this paper we present two microscopic models for quantum Hall plateau transitions.
The first model describes electrons hopping on a tight-binding lattice with uniform magnetic flux
and on-site disorder.
The second model describes electrons in continuum with a short-range impurity potential
projected into the lowest Landau level.
In this section we review the basics of the two models.
We also describe the Chern number calculation method used to extract
the localization length critical exponent, and the statistical measures used to categorize the data.

\subsection{The lattice model}
\label{subsec:lattice_model}
We consider the two-dimensional tight-binding model
\begin{equation}
\mathcal{H}=-\sum_{\langle i,j\rangle}\left (te^{i\theta_{ij}}c_i^\dagger c_j+h.c.\right)
+ \sum_i \epsilon_i c_i^\dagger c_i,
\label{eq:lattice_model}
\end{equation}
where $t$ is the hopping strength and we have $\theta_{ij}=\frac{e}{\hbar}\int_i^j A\cdot \mathrm{d}l$ in the presence of a
perpendicular magnetic field $B$.
We choose the Landau gauge $\vec{A}=(0,Bx,0)$.
The magnetic flux $\phi$ per unit cell is
\begin{equation}
\frac{\phi}{\phi_0}=\frac{Ba^2}{hc/e}=\frac{1}{2\pi}\sum_{\Box} \theta_{ij},
\end{equation}
where $\phi_0 = hc/e$ is the flux quantum, and the disorder potential $\epsilon_i$
are independent variables with identical uniform distribution on $[-W,W]$.
As the flux $\phi$ per unit cell varies, the clean model has a self-similar energy spectrum,
known as the Hofstadter butterfly.
We choose the flux $\phi$ per unit cell as $\phi_0 / 3$.
The clean Hamiltonian is translationally invariant and
can be diagonalized to three subbands,
\begin{equation}
H_0({\bf k}) |\phi_n({\bf k})\rangle=E_n({\bf k})|\phi_n({\bf k})\rangle,
\end{equation}
where the subband index $n = 0,1,2$.
The three subbands carry Chern number $1$,$-2$, and $1$,
and each contains $N_\phi=L_1 L_2/3$ states for an $L_1 \times L_2$ lattice.
In the full model the evolution of the conducting states in the lowest subband
is correlated with the central band and may introduce
additional complexity.~\cite{Yang1996}
Therefore, we truncate the Hilbert space by
projecting the disorder potentials into the lowest subband,
\begin{equation}
\tilde{V}=\sum_{{\bf k},{\bf k'}}|\phi_0({\bf k})\rangle\langle \phi_0({\bf k})|V({\bf r})|\phi_0({\bf k'})\rangle \langle \phi_0({\bf k'})|.
\end{equation}
Earlier work of Thouless conductance found that the critical behavior is independent
of the ratio $t/W$ when $t / W \leq 0.2$.~\cite{Wan2000}
Hence, we set $t = 0$ and $W = 1$ {\it after} the projection.
In this case the clean band width is reduced to zero
so we have particle-hole symmetry; this fixes the critical energy at $E_c = 0$.

\subsection{The continuum model}
\label{subsec:continuum_model}
 We consider the two-dimensional electron in presence of a magnetic field
\begin{equation}
H=\frac{1}{2m}(\mathbf{P}+\frac{e}{c}\mathbf{A})^2+V(\mathbf{r})=\frac{\mathbf{\Pi}^2}{2}+V(\mathbf{r}),
\end{equation}
where the symmetric gauge $\mathbf{A}=\frac{1}{2}B(y,-x)$ is used, and $V(\mathbf{r})$ is the
random potential. The generator of infinitesimal magnetic translation is
\begin{equation}
\kappa\equiv(\kappa_1, \kappa_2)
=\mathbf{P}-\frac{e}{c}\mathbf{A}=\mathbf{\Pi}(-B),
\end{equation}
with $[\kappa_1,\kappa_2]=-i$,
where we set the magnetic length $l_B = 1$ and $\hbar = 1$ for convenience.
For a finite $L_1\times L_2$ system,
the translation operator $t(L_j\hat{e}_j)=\exp(i\kappa_j L_j)$ satisfies the magnetic algebra
\begin{equation}
t({\bf a})t({\bf b})=\exp \left [ i({\bf a}\times {\bf b})\cdot \hat{z} \right ]t({\bf b})t({\bf a}).
\end{equation}
When the number of flux quanta in an $L_1 \times L_2$ sample
$N_\phi = L_1 L_2 / (2 \pi)$ is an integer,
$t(L_1\hat{e}_1)$ and $t(L_2\hat{e}_2)$ commute with each other.
By defining two primitive translations,
$t_j=t(L_j/N_\phi\hat{e}_j)=\exp(i\kappa_j L_j/N_\phi)$ for $j = 1, 2$,
we are able to construct a Landau-like stripe basis ${| \phi_m\rangle }$
in the lowest Landau level by requiring
\begin{equation}
\begin{cases}
t_1|\phi_m\rangle =|\phi_{m+1}\rangle,\\
t_2|\phi_m\rangle =\exp(-i2\pi m/N_\phi)|\phi_m\rangle. \\
\end{cases}
\end{equation}
We consider randomly placed scatterers with $\delta$-potential, i.e.
$V(\mathbf{r}) = \sum_i^{N_{\rm imp}} \delta(\mathbf{r} - \mathbf{r}_i)$,
whose Fourier components are $V_{\mathbf{q}}$.
The number of short-range scatterers is chosen to be $16 N_{\phi}$.
The random potential is then projected into the lowest Landau level
\begin{equation}
\tilde{V}=\sum_{j,k}|\phi_j\rangle\langle \phi_j|\sum_{\mathbf{q}}V_{\mathbf{q}}e^{i \mathbf{q} \cdot \mathbf{r}} |\phi_k\rangle \langle \phi_k|,
\label{eq:V(r)}
\end{equation}
where $\mathbf{q}=(q_1,q_2)=(2\pi m/L_1,2\pi n/L_2)$ and $m,n$ are integers.

For simplicity, we consider square samples $L_1 = L_2 = L$ with
generalized boundary condition
\begin{equation}
\label{eq:boundary_condition}
t(L\hat{e}_j) |\psi\rangle=e^{i\theta_j}|\psi\rangle,~j=1,2,
\end{equation}
with $\theta_j\in[0,2\pi]$,
the Hamiltonian matrix is calculated as
\begin{align}
H_{jk}(\theta_1,\theta_2)=&\sum_{m,n}V_{mn}e^{-\pi(m^2+n^2)/2N_\phi}e^{-i\pi mn/N_\phi} \nonumber \\
\times &e^{i[m\theta_2-n\theta_1]/N_\phi}e^{-i2\pi mj/N_\phi}\langle \phi_j|\phi_{k-n}\rangle.
\end{align}
For particle-hole symmetric potential, the critical energy is also expected at $E_c = 0$.

\subsection{The Chern number calculation}
\label{sec:Method}
In either case, we can impose generalized boundary conditions.
The boundary condition averaged Hall conductance for a state $\psi_m$
is a topological invariant~\cite{Niu1985}
\begin{equation}
\langle \sigma_{xy}(m)\rangle = C(m)\frac{e^2}{h},
\end{equation}
where the first Chern number
\begin{equation}
C(m) = \frac{1}{2\pi i}\int_0^{2\pi}\int_0^{2\pi}\mathrm{d}
\theta_x\mathrm{d}\theta_y
\left[\left\langle \frac{\partial \psi_m}{\partial \theta_x} \right .\left \vert \frac{\partial \psi_m}
{\partial \theta_y}\right\rangle -h.c.\right].
\end{equation}
For efficient numerical calculation, we follow the method of
Fukui et al.~\cite{Fukui08}
We divide the $2 \pi \times 2 \pi$ boundary condition space into
a grid of $L_g \times L_g$ sites
\begin{equation}
\vec{\theta}=(\theta_x,\theta_y)=\frac{2\pi}{L_g}(l_x,l_y), ~l_x,l_y=0,\cdots,L_g-1,
\end{equation}
so primitive vectors $\hat{\mu}$ are vectors in the directions of $\mu = x$ and $y$
with length $2 \pi / L_g$.
For the $m$th eigenstate, we define a $U(1)$ link variable
\begin{equation}
U_\mu^m(\vec{\theta})=\langle \phi_m(\vec{\theta})|\phi_m(\vec{\theta}+\hat{\mu})\rangle /N_\mu^m(\vec{\theta}),
\end{equation}
where $N_\mu^m(\vec{\theta})=|\langle\phi_m(\vec{\theta})|\phi_m(\vec{\theta}+\hat{\mu})\rangle|$
is the normalization factor.
The Berry phase in a unit cell at $\vec{\theta}$ is then calculated
by a gauge invariant formula
\begin{equation*}
\gamma_m(\vec{\theta})={1 \over i} \ln U^m_x(\vec{\theta})U^m_y(\vec{\theta}+\hat{x})
U^m_x(\vec{\theta}+\hat{y})^{-1}U^m_y(\vec{\theta})^{-1}.
\end{equation*}
For a sufficiently fine grid, the local Berry phase is small in amplitude and can be
restricted to $(-\pi, \pi]$ in numerical calculation.
The Chern number $C(m)$ is obtained by summing the phase over all lattice sites
\begin{equation}
\label{eq:discretesum}
C(m)=\frac{1}{2\pi}\sum_{\vec{\theta}} \gamma_m (\vec{\theta}).
\end{equation}
Due to the introduction of the locally gauge-invariant formulation
of Chern number, the efficiency of the method has been improved greatly even
with a coarsely discretized boundary condition space.
As pointed by Arovas et al.,\cite{Arovas1988}
the sensitivity of the nodes of the wave function to the smooth change
in boundary conditions can be used to distinguish the localized
and extended states.
Consequently, we follow Huo and Bhatt~\cite{Huo1992} to identify states
with nonzero Chern numbers as conducting states.

The total number of conducting states per sample is defined as
\begin{equation}
N_c=\int_{-\infty}^{\infty}\rho_c(E)\mathrm{d}E,
\end{equation}
where $\rho_c(E)$ is the density of conducting states.
In the vicinity of the critical energy $E_c$,
the localization length diverges as $\xi(E) \sim |E-E_c|^{-\nu}$.
For a finite size system with linear size $L\sim \sqrt{N_\phi}$,
one expects that the states with $\xi(E)>L$ are conducting.
The number of such states scale as
$N_c\sim L^2 \rho(E_c) |E_m-E_c| \sim N_\phi^{1-1/(2\nu)}$,
to the lowest order,
where $E_m$ is determined by $\xi(E_m) \sim L$.
One can also define the width of $\rho_c(E)$ as
the square root of its second moment
\begin{equation}
\Delta E=\left [ \frac{\int_{-\infty}^\infty(E-E_c)^2\rho_c(E)\mathrm{d}E}{\int_{-\infty}^\infty
\rho_c(E)\mathrm{d}E}\right ].
\end{equation}
Roughly speaking, the width is set by $L\sim \xi(E) \sim \Delta E^{-\nu}$,
so it is expected to follow $\Delta E \sim L^{-1/\nu} \sim N_\phi^{-1/(2\nu)}$.
The lowest-order scaling of $N_c$ and $\Delta E$
may need irrelevant length corrections.

\subsection{Goodness of fit}
\label{sec:DataAnalysis}

In this paper we compare various scaling hypotheses with and without
irrelevant length corrections.
After obtaining the best-fit values of parameters, we need
to decide which hypothesis is the best description of the data.
Our result of the critical exponent is then obtained
from the best-fit with the best hypothesis.
In this subsection, we discuss the acceptability of
the scaling hypothesis.

Generally speaking, we fit $N$ data points $(x_{n},y_{n})$ with $n=0,\dots,N-1$ with
measurement errors $\sigma_{n}$ to a model that has $M$ adjustable parameters
$a_{0},\dots,a_{M-1}$. The model has a functional relationship between
the measured variables and parameters $y(x)=y(x;a_{0}, \dots, a_{M-1})$,
minimizing the parameters $a_{0}, \dots, a_{M-1}$ is equivalent to
minimizing the chi-square
\begin{equation}
\chi^{2}\equiv\sum_{n=0}^{N-1}\left[\frac{y_{n}-y(x_{n};a_{0}, \dots, a_{M-1})}{\sigma_{n}}\right]^{2}.
\end{equation}
A typical value of $\chi^{2}$ for
a moderately good fit is $\chi^{2}\approx {\rm dof}$,
where ${\rm dof}=N-M$ is the degrees of freedom,
or, equivalently, the reduced chi-square $\bar{\chi}^{2}\equiv\chi^{2}/{\rm dof}\approx1$.
Relatively smaller value of $\chi^2$ indicates a better fit.
However, one must also take into account the number of adjustable parameters
used in the fit, to determine the overall acceptability of the fit.
A quantitative measure of the goodness-of-fit is given by the cumulative
probability function $Q(\chi^{2}|{\rm dof})$ , which
is the probability that the observed chi-square will exceed a particular
values $\chi^{2}$ by chance even for a correct model~\cite{NumericalRecipes}
\begin{equation}
Q(\chi^{2}|{\rm dof})=1-P\left (\frac{\rm dof}{2},\frac{\chi^{2}}{2} \right ),
\end{equation}
where $P(a,x)$ is the incomplete Gamma function.
A larger value of $Q$ is taken as an indicator of a better fit.
If $Q$ is too small, the reason could be either the model is unlikely to be
true and can be rejected, or the size of the measurement
errors $\sigma_{n}$ are larger than stated.
If $Q$ is close to $1$, it could be caused by overestimation of the
measurement errors.

\section{Results}
\label{sec:Results}
\subsection{The number of conducting states}
\label{subsec:nc_result}

\begin{figure}
\centering
\includegraphics[width=8cm]{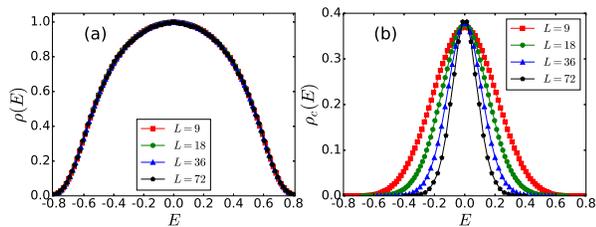}
\caption{(a) Disorder averaged total density of states $\rho(E)$ and (b) density of conducting
states $\rho_c(E)$ in the lattice model for system sizes $L=9$, 18, 36, and 72.}
\label{fig:dos}
\end{figure}

We begin with the study of square lattices with $L=6$-81.
The largest system contains almost 30 times more states than
the largest did twenty years ago.~\cite{Yang1996}
The number of disorder realizations ranges from $10^5$ for $L=6$
to $10^2$ for $L=81$.
In the Chern number calculation we choose a grid size
such that the number of grid points along in each dimension is
$L_g=30$ for $L\leq 30$ and $L_g\approx \sqrt{2/3}L$ for $L=36$-81.
Figure~\ref{fig:dos} shows the density of states $\rho(E)$ and
the density of conducting states $\rho_c(E)$
for several system sizes.

The total density of states $\rho(E)$ remains the same bell shape for all system sizes,
while the density of conducting states $\rho_c(E)$ shrinks
as the system size increases,
implying that in the thermodynamic limit only states with $E = 0$ are extended.
To prove this, we follow the same procedure in earlier literature~\cite{Huo1992,Yang1996}
to study the scaling behavior of the total number of conducting
states per sample $N_c$ and the width $\Delta E$ of the density of
conducting states $\rho_c(E)$, which are the zeroth and (the normalized) second moment of the density of nonzero Chern states.

\begin{figure}
\centering
\includegraphics[width=8cm]{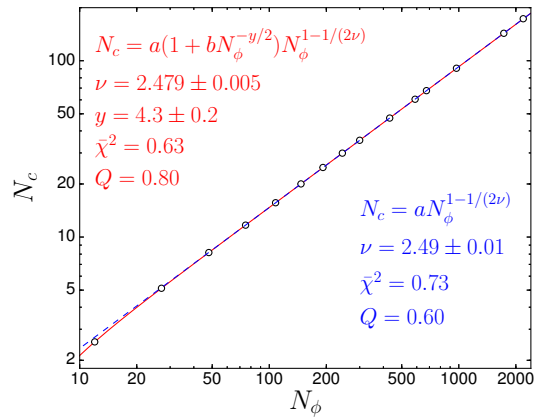}
\caption{Log-log plot of the number of conducting states per sample $N_c$
versus the total number of states per sample $N_\phi$ in the lattice model.
The percentage precisions for the data are 0.07\% to 0.2\% for $L=6$ to 54,
0.5\% for $L=72$, and 1\% for $L=81$.
The blue dashed line is the power-law fit without corrections (a straight line)
for $L \ge 30$.
The number of data, number of parameters, chi-squared, and goodness of fit
are 7, 2, 3.66, and 0.60, respectively.
The solid red line shows the fit with irrelevant length corrections for all the data.
The number of data, number of parameters, chi-squared, and goodness of fit
are 15, 4, 6.94, and 0.80, respectively.
}
\label{fig:lattice_Nc}
\end{figure}

We begin by fitting the number of conducting states to
\begin{equation}
\label{eq:powerfit}
N_c^{\rm power} = a N_\phi^{1-1/(2\nu)}.
\end{equation}
A chi-square fitting yields $\nu=2.49 \pm 0.01$,
as shown by the blue dashed line in Fig.~\ref{fig:lattice_Nc}.
The reduced chi-square $\bar{\chi}^2=0.732$ and
the goodness of fit $Q=0.599$.
The straightforward power-law fit for all system sizes results
in a good description of the data.
Deviation is visually noticeable only for $L = 6$
in Fig.~\ref{fig:lattice_Nc}.
This can be improved if we include the irrelevant length scale
corrections.

The leading correction can be accounted for by assuming
\begin{equation}
\label{eq:irrfit}
N_c^{\rm irr} = a(1+bN_\phi^{-y/2})N_\phi^{1-1/(2\nu)},
\end{equation}
where $y$ is the leading irrelevant length exponent.
A chi-square fitting now yields $\nu=2.479 \pm 0.005$
and $y=4.3 \pm 0.2$, as illustrated in Fig.~\ref{fig:lattice_Nc}.
The reduced chi-square $\bar{\chi}^2=0.631$ and the goodness of fit $Q=0.804$.
One expects this form is valid only for $|bN_\phi^{-y/2}|\ll1$.
Indeed, the correction term $bN_\phi^{-y/2}$ varies from $-0.06$ for $L=6$
to $-10^{-7}$ for $L=81$.
With or without the correction, the critical exponent $\nu$
remains unchanged within error bars.
This is supported by the similar $\bar{\chi}^2$ or $Q$ for the two fits.
The inclusion of the irrelevant perturbation, in principle, fits the data better
with the larger number of parameters.
However, it is debatable whether the improvement is enough
to warrant increasing the number of parameters.
It is worth pointing out that the best fit to the irrelevant length exponent is large,
$y = 4.3$. This is strong evidence that the irrelevant (or marginal)
perturbation at the criticality has a weak effect on the
Chern number calculation
and the scaling of the number of conducting states.

\begin{figure}
\centering
\includegraphics[width=8cm]{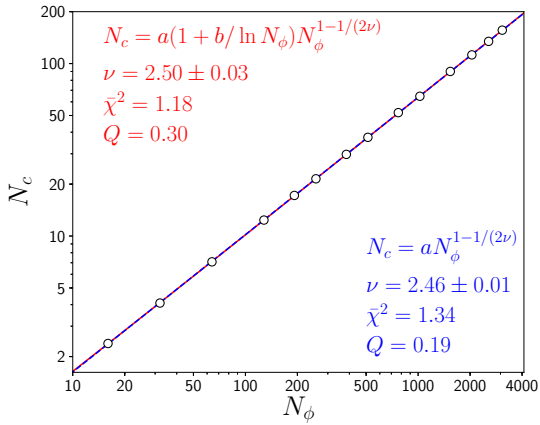}
\caption{Log-log plot of the number of conducting states $N_c$ vs. $N_\phi$
in the continuum model.
The percentage precisions of the data are 0.31\% to 0.48\%
for $N_\phi=16$ to 3,072.
The blue dashed line is the power-law fit
without any corrections.
The number of data, number of parameters, chi-squared, and goodness of fit
are 14, 2, 16.10, and 0.19, respectively.
The red solid line shows
the fit with logarithmic corrections.
The number of data, number of parameters, chi-squared, and goodness of fit
are 11, 3, 12.93, and 0.30, respectively.
}
\label{fig:continuum_Nc}
\end{figure}

For the continuum model, we study system size
from $N_{\phi}=16$ to 3,072, with disorder realizations
from 40,000 for $N_\phi=16$ to 608 for $N_\phi=$3,072.
In the Chern number calculation we use grid size $L_g=50$
for all system sizes.
Fig. \ref{fig:continuum_Nc} fits the number of conducting states as
$N_c=aN_\phi^{1-1/2\nu}$.
For all available 14 system sizes, we find $\nu=2.46 \pm 0.01$
with $\bar{\chi}^2=1.34$ and $Q=0.19$.
We emphasize that there are no irrelevant length corrections in
the fitting, and the exponent is in good agreement with the value
obtained above in the lattice model.
We also fit data only in systems with $N_\phi\geq 256$, the result is
$\nu=2.46 \pm 0.02$ with $\bar{\chi}^2=0.67$ and $Q=0.70$.
The exponent agrees with the previous one for all systems, but
the error bar is now larger due to the fewer system sizes.

The agreement in $\nu$ between all systems and larger system only
already suggests that the existence of the irrelevant length corrections
in the continuum model is difficult to demonstrate.
Attempt to fit all data with the corrections leads to large error bars in
$\nu = 2.47 \pm 0.01$ and $y = 8.7 \pm 23.6$,
which indicates that a single leading irrelevant length correction
is not a good hypothesis.
It is possible that the existence of a dangerous (or marginal) irrelevant operator may cause slower disappearing corrections to scaling.
We attempt to fit all data in the continuum model to a power law
with logarithmic corrections
\begin{equation}
\label{eq:logfit}
N_c^{\rm log} = a ( 1 + c / \ln N_{\phi} ) N_\phi^{1-1/(2\nu)}.
\end{equation}
As illustrated in Fig.~\ref{fig:continuum_Nc},
it yields $\nu = 2.50 \pm 0.03$ with $\bar{\chi}^2 = 1.18$
and $Q = 0.30$.
With the logarithmic corrections, the reduced chi-square is now smaller
while the goodness of fit is larger.
The resulting exponent $\nu = 2.50$ is larger but agrees, within the error bars, with the value in the absence of the corrections.

\begin{figure}
\centering
\includegraphics[width=8cm]{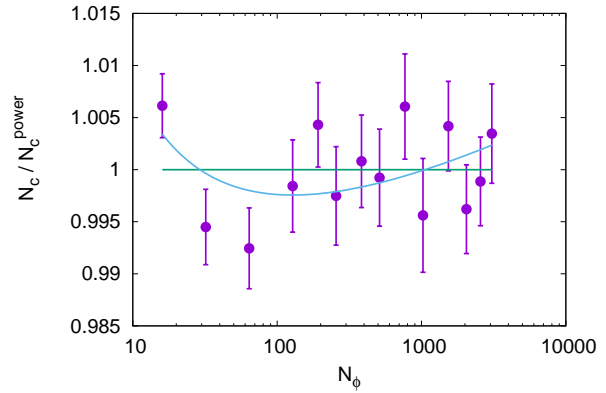}
\caption{Rescaled plot of Fig.~\ref{fig:continuum_Nc} of the power-law fits of the total number of conducting states without irrelevant length corrections (the horizontal line) and that with logarithmic corrections (the curved line). To visualize the errors of the data from very different sizes, we rescaled all data and fits by the power-law fit without corrections.
}
\label{fig:comparisons}
\end{figure}

We can compare the differences between the fit
without corrections [Eq.~(\ref{eq:powerfit})] and
the one with logarithmic corrections [Eq.~(\ref{eq:logfit})]
by plotting the ratio of the two, as well as the properly scaled raw data,
as shown in Fig.~\ref{fig:comparisons}.
In the figure the horizontal line at 1 represents
the power-law fit $N_c^{\rm power}$.
The scattered data points, scaled by $1/N_c^{\rm power}$, are all within 2$\sigma$, or twice the corresponding error bars.
The fit with log corrections $N_c^{\rm log}$, scaled by $1/N_c^{\rm power}$, curves up, fitting the data from the three smallest systems better than
the horizontal line.
Unless it is necessary to consider the systems as small as $N_{\phi}=16$
or 32, the power-law fit without corrections is essentially as good as
the one with logarithmic corrections.
But if so, it remains very interesting to understand why the
topological number calculation can support excellent scaling
for more than one order of magnitude
in linear scale starting from such small systems.
Based on the comparison, if the trend of data persists (including error bars), we would only be able to distinguish the two if we could approach $N_{\phi}>40$,000.

Combining the results of the different fits in both the lattice model and the continuum model, we conclude that the scaling of the total number of conducting states yields a consistent value $\nu = 2.48 \pm 0.02$.

\subsection{The width of the density of conducting states}
\label{subsec:dE_result}

\begin{figure}
\centering
\includegraphics[width=8cm]{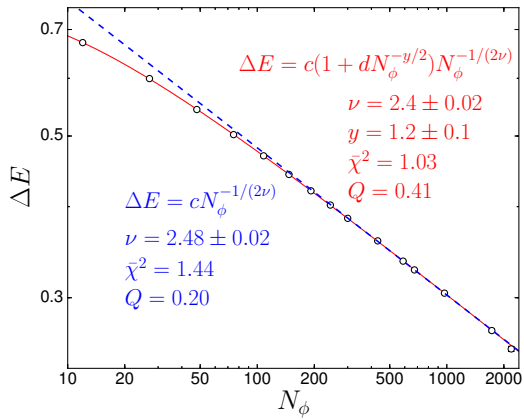}
\caption{Log-log plot of the width $\Delta E$ of the density of conducting states
vs. $N_\phi$ in the lattice model.
The percentage precisions of the data are  0.06\% to 0.17\% for $L=6$ to 54,
0.44\% for $L=72$, and 0.78\% for $L=81$.
The red solid line shows the power-law fit with leading irrelevant length corrections
for all data points.
The number of data, number of parameters, chi-squared, and goodness of fit
are 15, 4, 11.35, and 0.41, respectively.
The blue dashed line is the power-law fit without any corrections for systems
with $L\geq30$ (or $N_\phi\geq 300$).
The number of data, number of parameters, chi-squared, and goodness of fit
are 7, 2, 7.22, and 0.20, respectively.
}
\label{fig:lattice_dE}
\end{figure}

While the critical exponent $\nu$ is found to be universal in different models
as well as using different measures, as expected,
our fits including one irrelevant (or marginal) correction term
yields different results for the irrelevant length exponent from different quantities.
To show this, we consider the width $\Delta E$ of
the density of conducting states.
With the corrections of the leading irrelevant length,
one expects $\Delta E=c(1+dN_\phi^{-y/2})N_\phi^{-1/(2\nu)}$.
A chi-square fitting gives $\nu=2.40 \pm 0.02$ and $y=1.2 \pm 0.1$
with the reduced chi-square $\bar{\chi}^2=1.032$ and
the goodness of fit $Q=0.415$,
as shown in Fig.~\ref{fig:lattice_dE}.
The correction term $dN_\phi^{-y/2}$ varies from $-0.12$ for $L=6$ to
$-0.005$ for $L=81$.
The correction is twice as large for small system sizes as that in the
scaling of $N_c$, which suggests that we may need to consider additional corrections;
such a fit may result in a different value of $y$.

The comparison of the scalings of $N_c$ and $\Delta E$ shows that
even with the same set of data, namely $\rho_c(E)$,
one obtains exponent $\nu$ with a slight difference.
Judging from the smaller size of the corrections and, in particular,
the larger value of $y$, we conclude that the scaling of $N_c$ is more reliable.
This is consistent with empirical evidence from simulations of other disordered systems
such as spin glasses that lower order moments of distributions can be calculated more reliably (they also seem to converge faster to equilibrium than high order moments).
An additional evidence is that we can perform a power-law
fitting of $\Delta E$ without the irrelevant length corrections,
$\Delta E \sim c N_\phi^{-1/(2\nu)}$, for system sizes $L\geq 30$
(i.e., $N_\phi\geq 300$), as shown by the blue dashed line
in Fig.~\ref{fig:lattice_dE}.
The result is $\nu=2.48 \pm 0.02$ with $\bar{\chi}^2=1.444$ and $Q=0.205$.
Therefore, the value obtained from larger systems is consistent with
that obtained from the scaling of $N_c$.

The comparison suggests that the tails of the density of conducting states $\rho_c(E)$
suffer more from the finite-size effects.
The width $\Delta E$ is related to the second moment of $\rho_c(E)$,
while $N_c$ is the zeroth moment of $\rho_c(E)$.
Therefore, $\Delta E$ amplifies the finite-size fluctuations in the tails,
or at energies far from the critical energy.
If this understanding is correct, we could expect that in a different model but with
the same method the scaling of $N_c$ would generate the consistent
universal $\nu$, even though corrections to scaling could be different.
However, the scaling of $\Delta E$ using finite sizes could give a different $\nu$, whose value
depends on the non-universal finite-size fluctuations in the tails.

\begin{figure}
\centering
\includegraphics[width=8cm]{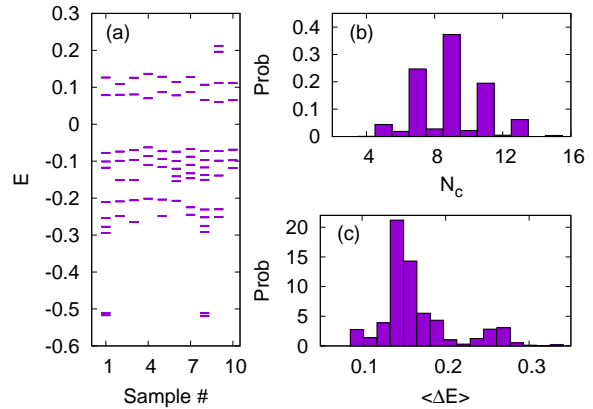}
\caption{(a) Energies of the conducting states in 10 samples ($L = 12$ or $N_{\phi} = 48$), whose on-site potentials differ only by 5\% fluctuations. Pairs of conducting states can emerge far from the band center. (b) Distribution of the number of conducting states in each sample from 1,000 realizations. With the constraint that the total Chern number per sample is $+1$, the distribution show strong odd-even difference, but is bell-shaped either for odd $N_c$ or for even $N_c$. (c) The root-mean-square energy for conducting states in each sample for the same 1,000 realizations. The distribution has a long tail with additional peaks.
}
\label{fig:tail}
\end{figure}

An independent confirmation of the tail effect comes from the study
of the perturbations to the disordered potential. We start with an arbitraty disorder realization in a 12$\times$12 lattice and perturb it with an additional disordered potential, whose strength is 5\% of the original one. The additional potential distorts the energy spectrum and likely causes a pair of
energy levels to cross each other. As a result, the Chern number of the pair can change. Since most states are localized, the change creates a pair of conducting states with Chern numbers $+1$ and $-1$, as demonstrated in Fig.~\ref{fig:tail}(a).
The pair, if lying in the tail, can affect the width of the conducting
states much more significantly than their total number.
For illustration, we analyze 1,000 samples which differ from each other only by the 5\% fluctuations.
Figure~\ref{fig:tail}(b) plots the distribution of the number of conducting states $N_c$ in each sample.
$N_c$ vary from 4 to 15 with strong odd-even fluctuations, because the total Chern number in each sample is constrainted to be $+1$ and most conducting states carry Chern number $+1$ or $-1$.
Other than the odd-even effect, the distribution is bell-shaped with a well defined peak at $N_c = 9$.
We expect that the total number is self-averaged in the thermodynamic limit and hence well-behaved.
On the other hand, Fig.~\ref{fig:tail}(c) plots the distribution of the root-mean-square energy of the conducting states in each sample for the same 1,000 realizations.
The distribution is clearly not bell-shaped.
Well above the main peak at 0.14, we can find a significant bump around 0.26 and the tail extends to as large as 0.33.
The multiple-peak structure is consistent with the emergence of conducting pairs far from the center, as shown in Fig.~\ref{fig:tail}(a).
But such pairs do not cause the distribution of $N_c$ to deviate significantly from the bell shape.
Therefore, We lean toward the scaling results of $N_c$, rather of $\Delta E$, based on the study of the fluctuations.

We study the scaling of the width $\Delta E$ of the
density of conducting states $\rho_c(E)$ in the continuum model next.
If we fit all data to the form $\Delta E=cN_\phi^{-1/(2\nu)}$,
we obtain $\nu=2.33 \pm 0.01$ with the reduced chi-square
$\bar{\chi}^2=1.50$ and the goodness of fit $Q=0.12$.
When we only fit data with $N_\phi\geq 256$,
we obtain $\nu=2.36 \pm 0.02$
with $\bar{\chi}^2=1.63$ and $Q=0.12$.
Within the error bars, we cannot draw a definite conclusion
on the trend of $\nu$ in the thermodynamic limit.
Like the scaling of $N_c$, including either power-law corrections
or logarithmic corrections shows no significant improvement.

\begin{figure}
\centering
\includegraphics[width=8cm]{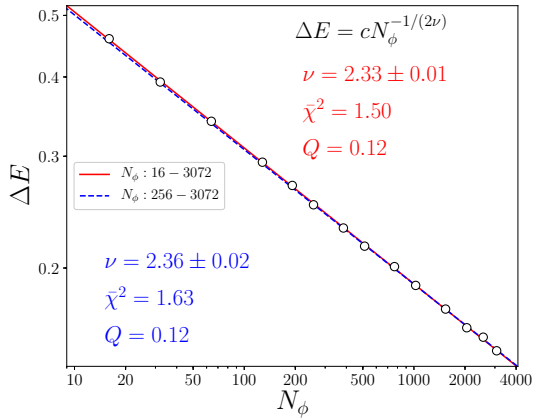}
\caption{Log-log plot of the width $\Delta E$ of the density of conducting states
vs. $N_\phi$ in the continuum model.
The percentage precisions of the data vary between 0.31\% and 0.45\% for
$N_{\phi}=16$ to 3,072.
The red solid
line is the power-law fit for the data points from $N_\phi=16$ to $N_\phi=3$,072.
The number of data, number of parameters, chi-squared, and goodness of fit
are 14, 2, 18.0, and 0.12, respectively.
The blue dashed line is the fit for the data points from $N_\phi=256$ to $N_\phi=2048$.
The number of data, number of parameters, chi-squared, and goodness of fit
are 9, 2, 11.4, and 0.12, respectively.
}
\label{fig:width_pwu}
\end{figure}

As we discussed at the end of Sec.~\ref{subsec:nc_result},
the localization length critical exponent $\nu$ obtained from the scaling of
$N_c$ is consistent in the lattice model and the continuum model.
However, the values of $\nu$ obtained from the scaling of $\Delta E$
in the two models are not.
The observations confirm that the finite-size fluctuations are
more significant in the tails of the density of conducting states.
In other words, the values obtained from the scaling of $\Delta E$
are less reliable, unless we can approach much larger system sizes.
We also studied the width of the density of conducting states for different percentiles and found that the fluctuations are too large to allow accurate analysis.

\subsection{Errors in the Chern Number calculation}
\label{subsec:errors}

Because Chern number, or the dimensionless Hall conductance,
is a topological invariant,
the main errors in the Chern number approach come from the
discretization of the boundary condition space into a grid.
The accuracy of Chern numbers thus depends on
the grid size $L_g$.
But increasing $L_g$ will slow down the Chern number calculation.
To estimate the optimal $L_g$, e.g., in the lattice model,
we consider uniformly distributed $\gamma_m$ in Eq.~(\ref{eq:discretesum})
in the boundary condition space.
For a state $m$ with Chern number $C(m)$, we expect
\begin{equation}
\bar{\gamma}_m = \frac{2 \pi C(m)}{L_g^2},
\end{equation}
for each $\vec{\theta}$.
Sign errors of $\gamma_m$ occur due to the restriction of the phase
to $(-\pi,\pi]$ when $\gamma_m$ is comparable to $\pi$, i.e.,
\begin{equation}
\frac{2 \pi C(m)} {L_g^2} \sim \epsilon \pi
\end{equation}
where $\epsilon$ can be understood as the error rate of the state.
This gives a rough estimate of the error rate per state with
$C(m) = O(1)$ as $\epsilon \sim 2 / L_g^2$.

We implement a simple consistency check for the Chern number calculation:
the total Chern number of each sample should be one.
When we obtain a sample that fails the total Chern number check,
we reject all states in the sample.
This happens when at least one of the states in the sample has an
incorrect Chern number.
For small enough $\epsilon$, the rejection rate $r$ for a sample with
$L^2/3$ states is
\begin{equation}
\label{eq:r_rate}
r \approx \epsilon \left (L^2/3 \right ) \sim {2 \over 3}  \left (L \over L_g \right )^2
\end{equation}
Therefore, in the lattice model we choose $L_g \approx \sqrt{2/3} L$
for $L \geq 36$ and $L_g = 30$ for $L \leq 30$,
in order to keep the rejection rate low.

We also run numerical tests to estimate the error rate of the
Chern number calculation for individual states and
the consequent errors in $N_c$ and $\Delta E$,
which we use to extract $\nu$.
In the tests we study the computational errors of lattice systems
with linear size $L = 6$-21, each with $N_s =10,000$
random potential realizations.
Our calculation finds that the rejection rate $r$ vanishes
as $L_g^{-\eta}$ with $\eta$ varying between 2.5-3.3,
which decreases even faster than our simple expectation.
When we fix $L_g = 30$, the rejection rate $r$ is no more than
$5 \times 10^{-3}$ and increases with size as
$L^{1.6}$, which is close to $L^2$ as we expect in Eq.~(\ref{eq:r_rate}).
The better-than-expected observations are likely due to the fact that
the majority of the Chern numbers is zero and their percentage
increases with $L$.
With the numerical results, we can assume that Chern numbers
calculated with $L_g = 60$ are sufficiently accurate and can be used as a reference.
We then calculate Chern numbers with $L_g = 30$.
By comparing them, we estimate the relative error
in $N_c$ is around $10^{-3}$ and
the relative error in $\Delta E$ is even smaller.
Compared to the sample-to-sample fluctuations, we find
the errors due to the Chern number calculation are at least
one order of magnitude smaller and can be neglected.
The measurement errors in the preceding subsections are,
therefore, the standard error in the mean of the relevant data.

\section{Summary and Discussion}
\label{sec:conclusions}

In this work, we have conducted a detailed study of
two models for the integer quantum Hall plateau transition:
the disordered Hofstadter lattice model and the continuum Landau-level model.
We perform high-precision Chern number calculation for system size
up to $N_{\phi} = 2$,187 in the lattice model and 3,072 in the continuum model.
We use multiple measures to rule out the errors in the Chern number calculation caused by
the discretization of the integration grid.
Using nonzero Chern number as criterion, we calculate
the density of conducting states $\rho_c(E)$ for various system
sizes.

By fitting the total number of conducting states $N_c$
to a power law, we obtain the localization length critical exponent
$\nu = 2.49 \pm 0.01$ in the lattice model and
$\nu = 2.46 \pm 0.01$ in the continuum model.
In the lattice model we obtain $\nu = 2.479 \pm 0.005$
with better goodness of fit
after we include the leading irrelevant length corrections.
However, the bare power-law fit cannot be improved in the continuum model
by the inclusion of the leading irrelevant length corrections.
With systems as small as $N_{\phi} = 16$, a fit with logarithmic corrections, which may be caused by a dangerous (marginal) irrelevant perturbation, yields $\nu = 2.50 \pm 0.03$.
Based on these results, we conclude that the two models are in the same universality class and $\nu = 2.48 \pm 0.02$.
The larger exponent is found in much larger systems than
in the earlier studies.
In fact the largest system in the present study has 24 times
the flux quanta of that
in the earlier study for the continuum model,~\cite{Huo1992}
and almost 30 times the flux quanta for the previous lattice model.~\cite{Yang1996}
However, given the fact that earlier studies had large error bars of
0.1, the new result is still consistent with the earlier
estimates.~\cite{Huo1992,Yang1996,bhatt02}

Our analysis of both models, using fully two-dimensional scaling,
suggests that corrections to scaling due to irrelevant operators are small,
unlike what is found in strip-geometry methods
using the crossover to one-dimension to extract critical exponents.
Even in the lattice model where the corrections are apparently larger,
the leading irrelevant length exponent is found to be
$y = 4.3 \pm 0.2$, which is substantially larger than the value
found in the methods using the crossover to one dimension using
the strip geometry on the CCN model.
This suggests that in the topological number based calculation
the leading irrelevant correction may have small amplitude.
Because of this, the precise value of the best-fit irrelevant length exponent
varies depending on the model and presumably therefore has greater uncertainty,
though it is certainly large.
On the other hand, the smaller corrections to scaling would explain
why the earlier Chern number calculations
revealed the consistent result of $\nu = 2.4 \pm 0.1$
in systems with no more than 128 magnetic flux quanta.~\cite{Huo1992}

We also attempt to determine $\nu$ from the scaling of the width
of the density of conducting states $\rho_c(E)$,
which is related to the second moment of $\rho_c$.
The value is found to be appreciably smaller than that
obtained by the scaling of $N_c$ and is model-dependent:
$2.40 \pm 0.02$ for the lattice model and
$2.33 \pm 0.01$ for the continuum model.
Fitting the data from larger systems only in the lattice model
gives $\nu = 2.48 \pm 0.02$, which is then consistent with the value
from the scaling of $N_c$.
This suggests that higher moments (e.g. the second moment) of
the distribution of the nonzero Chern number states are subject to
greater error than the total number (zeroth moment).
Our further analysis attributes this to the random flutuations of
conducting states in the band tails.
Therefore, the agreement of the critical exponent obtained from
the total number and the width of the conducting states
in the earlier study~\cite{Huo1992} are likely due to the large error bars.

The localization length critical exponent $\nu = 2.48 \pm 0.02$
in the present study differs from the value (varying from 2.55 to 2.62)
found in recent studies of the CCN
model.~\cite{Slevin2009,Obuse2010,Amado2011,Dahlhaus2011,Fulga2011,Slevin2012,Obuse2012}
One possibility, which requires further work, is the issue of whether the CCN model and the non-interacting quantum Hall transition in a Landau level
belong to the same universality class, as was believed for several years.
This has recently been challenged by Gruzberg and coworkers,~\cite{gruzberg17}
who argued that the CCN model may not capture all types of disorder that are relevant at the integer quantum Hall plateau transition.
However, the authors~\cite{gruzberg17} found in the numerical simulation of
a geometrically disordered network model $\nu = 2.374 \pm 0.018$, which is
significantly smaller than that found in the present study for
the lattice model and the continuum model.

Very recently, {\em after our manuscript was in the referee process}, Puschmann and coworkers~\cite{puschmann18} found a localization length exponent $\nu = 2.58 \pm 0.03$ in strips of microscopic lattices, which is consistent with the CCN model.
The lattice model contains all subbands, unlike the one used in the present study, which projects the disordered potential to the lowest subband with
Chern number $+1$.
The authors~\cite{puschmann18}
employed a scheme based on crossover to one-dimensional systems and
showed that the scaling of the Lyapunov exponent depends on flux per plaquette $\phi$ and converges for $\phi \leq 1/10$.
In particular, they found that the corrections to scaling for $\phi = 1/3$ and $1/4$ deviate significantly from the small flux limit, which are expected to be identical to the results of the continuum model.
In contrast, we find that in the projected lattice model, in which there are no subband mixings by design, the results for $\phi = 1/3$ is consistent with those in the continuum model, {\em with or without corrections}, based on  purely two-dimensional scaling analysis.
We emphasize that in both lattice studies, the scaling behavior persists for well over one order of magnitude in linear size change.

There is yet another possibility for the remaining discrepencies. Zirnbauer~\cite{zirnbauer18} suggested that $1/\nu$ and $y$ can keep decreasing as the renormalization group fixed point is approached.
While we find no direct evidences supporting the scenario, this means that the approach to the ultimate scaling behavior can be much slower than we thought.
If this is the case, the puzzle on the localization length exponent can only be resolved with much larger system sizes.

\section{Acknowledgements}
\label{sec:acknowledgements}
The authors thank Ferdinand Evers, Ilya Gruzberg, Tomi Ohtsuki, Martin Puschmann, Keith Slevin, and Thomas Vojta for helpful discussions.
This work was supported by the 973 Program under Project No.
2012CB927404, by the NSFC under Grant No. 11174246,
and by U.S. DOE-BES Grant {DE-SC}0002140 (RNB).
RNB also thanks the hospitality of the Aspen Center for Physics for a stay
during which the manuscript was being finalized.

\appendix
\section{Universality from the Lattice to the Continuum Model}
\label{app:moreLattice}

In the main text we study the lattice model and the continuum model, and our results confirm that the two are in the same universality class, as was expected in earlier studies.~\cite{huckestein95}
The particular choice of magnetic flux per plaquette we made is $\phi = 1/3$.
In the pure case, the band width of the lowest magnetic subband is not small (in units of the hopping strength $t$), compared with the gap separating the lowest and the central subbands.
To avoid band mixing, we project the Hamiltonian to the subspace of the lowest subband.
A side advantage is that we deal with a smaller Hilbert space dimension, hence the Chern number calculation is faster.
The choice of $\phi$ is not restricted.
In the case of $\phi = 1 / i$, where $i > 2$ is an integer, the spectrum of the full Himiltonian contains $i$ subbands (the central two are touching at $E = 0$ for even $i$).
The lowest (and the highest) subband for every integer $i$ is characterized by Chern number $+1$ and can be used to study the plateau transition.
For sufficiently large $i$, the band width of the lowest subband in the pure case is much smaller than the adjacent band gap, hence the spectral projection is not necessary.
In principle, other subbands, with the exception of the central band(s),
can model a Landau level as well; but they are known to have longer localization length, which can be overcome by, e.g., a correlated disorder potential.~\cite{Wan2000}

In addition to the localization length $\xi$ and the microscopic lattice constant $a$, we have a third length scale: the magnetic length
$l_B = a \sqrt{i / (2\pi)}$.
In the limit of large $i$, $a/l_B$ goes to zero.
Therefore, we can regard the continnum model as the limit of large $i$
and expect that the localization length critical exponent extracted in the lattice model with larger $i$ is consistent with the results for $\phi = 1/3$
and in the continnum model.
However, {\em subsequent to the online publication of this work}, a study of the
tight-binding lattice model in the absence of the subband projection~\cite{puschmann18} found $\nu = 2.58 \pm 0.03$ for $\phi \geq 1/10$
with pronounced corrections to scaling.
The authors there applied the Green's function method on a strip geometry and perform finite-size scaling of the dimensionless Lyapunov exponent by varying
the strip width.
For $\phi = 1/3$, $1/4$, and $1/5$, the authors found non-universal behavior,
which they attributed to the non-negligible intrinsic LL width.

\begin{figure}[t]
\centering
\includegraphics[width=8cm]{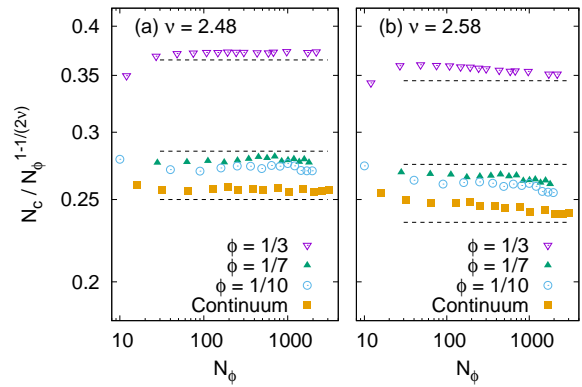}
\caption{System-size dependence of $N_c / N_{\phi}^{1 - 1/(2\nu)}$
for $\phi = 1/3$, $1/7$, $1/10$, and the continuum model under
assumption of (a) $\nu = 2.48$ and (b) $\nu = 2.58$ for the localization length exponent $\nu$.
Horizontal dashed lines are guides to the eye only.
}
\label{fig:evolution}
\end{figure}

In this Appendix, we provide additional data and analysis to show that the critical exponent $\nu = 2.48 \pm 0.02$ in the main text is universal for both the projected lattice model, regardless of $i$, and the continuum model.
In the projected model, the width of the magnetic subband (or the intrinsic LL width) is set to zero.
Figure~\ref{fig:evolution} shows the dependence of $N_c / N_{\phi}^{1 - 1/(2\nu)}$ on system size for $\phi = 1/3$, $1/7$, $1/10$, and in the continuum model.
The ratio is expected to be a horizontal line if $\nu$ is properly chosen and if there is no need for irrelevant length scale corrections.
We compare the assumptions of $\nu = 2.48$ (supported by the current study) and 2.58 (supported by a study of the lattice model without subband projection~\cite{puschmann18}).
Under the assumption of $\nu = 2.48$ as shown in Fig.~\ref{fig:evolution}(a),
the ratio is essentially a constant for $\phi = 1/3$ and the continuum model for almost two orders of magnitude in $N_{\phi}$.
For $\phi = 1/7$ and $1/10$, the ratio stays convincingly as a constant up to the largest systems we can access using Chern number calculation with meaningful sample average.
The relative uncertainties are about twice larger than the $\phi = 1/3$ case.
The difference between the four cases lies in the finite-size deviation of the data.
For $\phi = 1/3$, the ratio lies below the larger-system value for $L = 6$, or $N_{\phi} = 12$.
For other cases of the lattice model and for the continuum model, the ratio for the smallest systems (not shown here for $\phi = 1/7$) lies above the corresponding larger-system value.
The finite-size artifacts, however, do not affect the universality of the localization length exponent in these cases, according to Fig.~\ref{fig:evolution}(a).
On the other hand, Fig.~\ref{fig:evolution}(b) tests the power-law scaling under the assumption $\nu = 2.58$.
In sharp contrast to the results under $\nu = 2.48$, the larger exponent leads to clearly identifiable downward trend of the system-size dependence, regardless of the direction of the deviations in small systems.
In fact, the best power-law fits without any corrections yield $\nu = 2.50 \pm 0.01$ for $\phi = 1/7$ ($N_{\phi} \geq 28$) and $\nu = 2.50 \pm 0.02$ for $\phi = 1/10$ ($N_{\phi} \geq 40$), both consistent with $\nu = 2.48 \pm 0.02$ within error bars.

Based on the comparison, we conclude that, at least in the projected lattice model, the power-law scaling with $\nu = 2.48$ describes the total number of conducting states well {\em without any need of corrections and regardless of the value of $\phi$}.
This seems to be the most natural conclusion, given that the same value is also supported by the continuum model, which can be regarded as the limit of vanishing $\phi$.
Unless extremely slowly developing corrections exist, the Chern number calculations of the projected lattice model and the continuum model do not support $\nu = 2.58$ or any value close.


%

\end{document}